\input{aipcheck}

\documentclass[
    ,final            
  ]
  {aipproc}

\layoutstyle{6x9}

\begin{document}

\title{Compton reflection in AGN with Simbol-X}


\author{V. Beckmann}{
  address={ISDC Data Centre for Astrophysics, Ch. d'Ecogia 16, 1290
  Versoix, Switzerland}
}

\author{T.J.-L. Courvoisier}{
 address={ISDC Data Centre for Astrophysics, Ch. d'Ecogia 16, 1290
 Versoix, Switzerland}
}

\author{N. Gehrels}{
  address={NASA Goddard Space Flight Center, Code 661, Greenbelt, MD 20771, USA}
}

\author{P. Lubi\'nski}{
 address={Centrum Astronomiczne im. M. Kopernika, Bartycka 18,
PL-00-716 Warszawa, Poland}
}

\author{J. Malzac}{
  address={CESR, OMP, UPS, CNRS;
   B.P. 44346, 31028 Toulouse Cedex 4, France}
}

\author{P.~O. Petrucci}{
  address={Laboratoire d'Astrophysique de Grenoble, CNRS, UMR
   5571, BP 53X, 38041 Grenoble, France}
}

\author{C.~R. Shrader}{
  address={NASA Goddard Space Flight Center, Code 661, Greenbelt, MD 20771, USA}
}

\author{S. Soldi}{
  address={ISDC Data Centre for Astrophysics, Ch. d'Ecogia 16, 1290
 Versoix, Switzerland}
}

\begin{abstract}
AGN exhibit complex hard X-ray spectra. Our current understanding is
that the emission is dominated by inverse Compton processes which take
place in the corona above the accretion disk, and that absorption and
reflection in a distant absorber play a major role. These processes
can be directly observed through the shape of the continuum, the
Compton reflection hump around 30 keV, and the iron fluorescence line
at 6.4 keV. We demonstrate the capabilities of Simbol-X to
constrain complex models for cases
like MCG-05-23-016, NGC 4151, NGC 2110, and NGC 4051 in short (10 ksec) observations. We compare the
simulations with recent observations on these sources by INTEGRAL,
Swift and Suzaku. Constraining reflection models for AGN with Simbol-X
will help us to get a clear view of the processes and geometry near to
the central engine in AGN, and will give insight to which sources are
responsible for the Cosmic X-ray background at energies $>20 \rm \, keV$.

\end{abstract}

\maketitle


\section{Why is Compton-reflection in AGN of importance?}

The extragalactic X-ray sky is dominated by active galactic nuclei
(AGN) and clusters of galaxies. Studying the population of sources in
X-ray bands has been a challenge ever since the first observations by
rocket borne X-ray detectors. Above 2 keV a synopsis of previous results is as follows: the 2 -- 10 keV Seyfert~1 continua 
are approximated by a $\Gamma \simeq 1.9$ powerlaw form \cite{Sy1average}. A flattening 
above $\sim 10 \rm \, keV$ has been noted, and is commonly attributed to Compton reflection \cite{george}. 
There is a great deal of additional detail in this spectral domain: ``warm'' absorption, 
multiple-velocity component outflows, and relativistic line broadening. The Seyfert~2 objects are more poorly categorized here, but the general 
belief is that they are intrinsically equivalent to the Seyfert~1s, but viewed through much larger absorption columns. 

Above 20 keV the empirical picture is less clear. The $\sim 20 - 200
\rm \, keV$ continuum shape of both Seyfert types is consistent with a
thermal Comptonization spectral form, although in all but a few cases the data are not sufficiently constraining to rule out 
a pure powerlaw form with photon index $\Gamma \simeq 2.0$ for
Seyfert~1 and $\Gamma \simeq 1.8$ for Seyfert~2 \cite{AGN2nd}. Nonetheless, the non-thermal
scenarios with pure powerlaw continua extending to $\sim \rm \, MeV$
energies reported in the pre-{\it CGRO} era are no longer widely
believed, and are likely a result of background systematics. However,
a detailed picture of the Comptonizing plasma - its spatial,
dynamical, and thermo-dynamic structure - is not known. Among the
critical determinations which {\it Simbol-X} can provide are the
plasma temperature and optical depth (or Compton ``Y'' parameter) for
a large sample of objects. In order to understand the physical
processes and the location of the processing regions with respect to
the central black hole in AGN, a decoupling of the components is
essential. {\it Simbol-X} will help to determine exact measurements of
the Compton reflection strength together with the continuum shape and
the strength and width of the iron K$\alpha$ fluorescence line.

Related to the compilation of AGN surveys in the hard X-rays is the
question of what sources form the cosmic X-ray background (CXB). 
The most
reliable measurement in the 10 - 500 keV has been provided by 
{\it HEAO 1} A-4, showing that the CXB peaks at an energy of about $30 \rm \, keV$ \cite{Gruber}. The isotropic nature of the X-ray background points to an extragalactic origin, and as the brightest persistent sources are AGNs, it was suggested early on that those objects are the main source of the CXB \cite{Setti}. In the soft X-rays this concept has been proven to be correct through the observations of the {\it ROSAT} deep X-ray surveys, which showed that $90 \%$ of the $0.5 - 2.0 \rm \, keV$ CXB can be resolved into AGNs \cite{ROSATdeep}. At higher energies ($2 - 10 \rm \, keV$), {\it ASCA} and {\it Chandra} surveys measured the hard X-ray luminosity function (XLF) of AGNs and its cosmological evolution. These studies show that in this energy range the CXB can be explained by AGNs, but with a higher fraction of absorbed ($N_H > 10^{22} \rm \, cm^{-2}$) objects than in the soft X-rays (e.g. \cite{Ueda03}). 
A study based on the {\it RXTE} survey by Sazonov \& Revnivtsev
\cite{RXTENGC} derived the local hard X-ray luminosity function of
AGNs in the 3--20 keV band. They showed that the summed emissivity of
AGNs in this energy range is smaller than the total X-ray volume
emissivity in the local Universe, and suggested that a comparable
X-ray flux may be produced together by lower luminosity AGNs,
non-active galaxies and clusters of galaxies. Using the {\it HEAO
  1}-A2 AGNs, Shinozaki et al. (2006), however, obtained a local AGN
emissivity which is about twice larger than the value of Sazonov \&
Revnivtsev \cite{RXTENGC}. 

{\it INTEGRAL} and {\it Swift} added
substantial information to the nature of bright AGNs in the hard
X-rays in the local
Universe. Considering the expected composition of the hard X-ray
background, it does not currently appear that the population detected
by these missions can explain the peak at 30 keV, as Compton thick
AGNs are apparently less abundant than expected
\cite{INTXLF,AGNunification}. The fraction of Compton thick AGNs 
is found to be small ($< 5\%$, \cite{AGN2nd}). Evolution of the
source population can play a major role in the sense that the fraction
of absorbed sources among AGNs might be correlated with redshift \cite{Worsley}. {\it INTEGRAL} and
{\it Swift} probe the AGN population in the local Universe ($z <
0.1$). Because of this {\it INTEGRAL}/IBIS and {\it Swift}/BAT surveys
will most likely not be able to test evolutionary scenarios of AGNs and thus will be inadequate to explain the cosmic X-ray background at $E > 20 \rm \, keV$. 
Future missions with focusing optics such as {\it Simbol-X} 
are required to answer the question of what dominates the Universe in
the hard X-rays.




\section{Measurements compared to Simbol-X simulations}

\begin{figure}
  \includegraphics[width=7cm,angle=270]{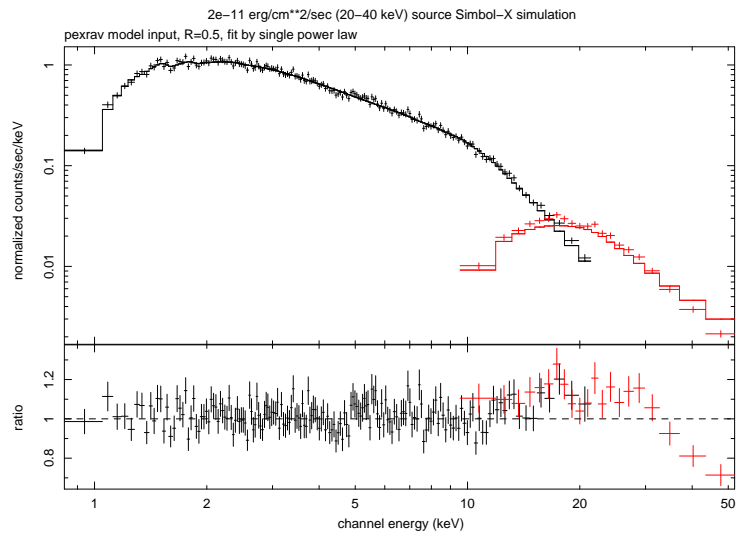}
  \caption{Simbol-X simulation. The simulated spectrum with a
    reflection strength of $R = 0.5$ is fitted by a single power
    law model. The residuals above 15 keV show the effect of the missing reflection
    component in the model.}
\label{fig:NGC4051}
\end{figure}

The strength of the Compton reflection
component has been measured using hard X-ray data from various
missions, such as {\it BeppoSAX}, {\it INTEGRAL}, {\it Suzaku}, often
in combination with soft X-ray measurements from {\it XMM-Newton},
{\it Chandra}, and {\it Swift}/XRT. The reflection strength $R$ is defined
as the relative amount of reflection compared to the directly viewed
primary spectrum. For some objects relative reflection values of $R >
1$ have been reported. 
This implies that more primary X-ray radiation is emitted 
toward the reflector than toward the observer due to e.g. relativistic effects 
caused by a dynamic corona moving towards the reflecting disk or light bending effects, or a special geometry with a high intrinsic covering fraction of the cold disk material. This can also be explained by variable emission of the central engine and a time delay between the underlying continuum and the 
reflected component from remote material.
In the following and in Tab.~\ref{tab:sources}, we compare the
measured reflection strength with what can be expected in 10 ksec
observations by {\it Simbol-X}:

\begin{table}
\begin{tabular}{lrrrr}
\hline
  & \tablehead{1}{r}{b}{NGC 4151}
  & \tablehead{1}{r}{b}{NGC 2110}
  & \tablehead{1}{r}{b}{MCG-5-23-16}
  & \tablehead{1}{r}{b}{NGC 4051}\\
\hline
$f_{20 - 40 \rm \, keV}\, [\rm erg \, cm^{-2} \, s^{-1}]$ & $24 \times 10^{-11}$ & $7.4 \times 10^{-11}$  &  $6 \times 10^{-11}$   & $2 \times 10^{-11}$\\
{\it INTEGRAL} exposure & 400 ks & 160 ks & 310 ks  & 700 ks\\
reflection ({\it INTEGRAL}) & $R = 1.0 {+0.4 \atop -0.3}$
  \cite{NGC4151} & $R < 0.1$ & $R<0.25$ \cite{MCG}  &
  $R = 5.6$\\
Other observatory & {\it BeppoSAX} + {\it XMM} & {\it BeppoSAX} & {\it
  Suzaku}  \cite{Reeves07}& {\it Suzaku} \\
reflection    & $R \sim 2$ \cite{Schurch} & $R<0.4$ \cite{Risaliti02}
  & $R = 1.1 \pm 0.2$ & $R \sim 7$ \cite{Terashima09}\\
Simbol-X simulation input& $R = 1.0$ & $R=0$ &  $R=0$ &  $R=0.5$\\ 
Simbol-X output&$R = 1.0 \pm 0.1$ & $R<0.04$ &  $R<0.1$ &  $R=0.6 \pm 0.2$\\ 

\hline
\end{tabular}
\caption{Reflection measurements for 4 different Seyfert galaxies,
  based on {\it INTEGRAL} and other missions, compared with {\it Simbol-X} simulations}
\label{tab:sources}
\end{table}
\begin{description}
\item[NGC 4151] This source is among the brightest AGN in the hard
  X-ray sky. Based on a 400 ks observation, {\it INTEGRAL} data
  allowed to constrain the reflection strength to $R = 1.0 {+0.4 \atop
    -0.3}$ \cite{NGC4151}, consistent with an earlier measurement based on {\it
    XMM-Newton} and {\it BeppoSAX} data ($R \sim 2$, \cite{Schurch}). A 10 ks Simbol-X
  observation would allow to constrain the value to 10\%. 
\item[NGC 2110] {\it BeppoSAX} observations constrained the reflection
  component to $R<0.4$ \cite{Risaliti02}, and recent simultaneous {\it INTEGRAL}/IBIS and
  {\it Swift}/XRT data resulted in $R<0.1$. A {\it Simbol-X}
  simulation, assuming $R=0$ resulted in an upper limit of $R<0.04$.
\item[MCG-5-23-16] {\it INTEGRAL} combined with simultaneous {\it
    Swift}/XRT data gave an upper limit of $R<0.25$ \cite{MCG}, while a 220 ks
  {\it Suzaku} observation derived $R= 1.1 \pm 0.2$ \cite{Reeves07}, leading to the
  conclusion that the reflection component is variable in this
  source. If indeed $R=0$, {\it Simbol-X} will be able to constrain
  this value to $R<0.1$ within 10 ks.
\item[NGC 4051] This source exhibits the strongest reflection
  component measured so far, with $R = 5.6$ (700 ks {\it INTEGRAL}) and
  $R\simeq 7$ (80 ks {\it Suzaku}, \cite{Terashima09}). A source at the flux level of NGC
  4051 but with a reflection component $R=0.5$ can be constrained to
  $R= 0.6 \pm 0.2$ in a 10 ks {\it Simbol-X} observation (Fig.~\ref{fig:NGC4051}).
\end{description}

\section{Conclusions}

While the reflection strength is already rather well constrained in
  a few X-ray bright AGN with $f_{20 - 40 \rm
  \, keV} > 5 \times 10^{-11} \rm \, erg \, cm^{-2} \, s^{-1}$ by data from
  existing missions, focusing optics as provided by {\it Simbol-X} are
  necessary in order to significantly increase our knowledge about how
  frequent strong reflection components are in AGN in general. In order to answer
  the question, whether the known AGN population in the local universe
  is sufficient to explain the CXB, or whether evolution has to be
  taken into account, {\it Simbol-X} will reveal whether in the
  majority of AGN $R \simeq 0$, $R \simeq 1$, or $R \gg 1$. This will
  also answer the question, how important reflection really is in the
  view of the central engine. 
10 ks
  observations are sufficient to constrain $R$ to $10
  - 20\%$ at the $2 \times 10^{-11} \rm \, erg \, cm^{-2} \, s^{-1}$ flux
  level. The second {\it INTEGRAL} AGN catalogue \cite{AGN2nd} includes 55 AGN above
  this flux level. These
  simulations show also that with {\it Simbol-X} the study of the reflection
  component simultaneously with the iron line will be possible for the
  first time on a dynamical time scale. This will provide crucial
  information for our understanding of the reflection component itself
  and of the AGN environment in general.


\end{document}